\documentclass[11pt]{article}
\usepackage{latexsym}
\usepackage{amssymb}
\usepackage{amsmath}
\usepackage{cases}
\usepackage{amsfonts}
\usepackage[dvipdf]{graphicx}
\usepackage{psfrag}

\makeatletter
 \@addtoreset{equation}{section}
 
 \makeatother
\begin{document}

\title{Probabilistic Uncorrelated Cloning Requires Negative Probability}

\author{Yuji Sekino\\
{\it Graduate School of Integrated Arts and Sciences, 
 }\\{\it Hiroshima University, Higashi-Hiroshima, 739-8521, Japan}}


\date{\today}
\maketitle

\begin{abstract}
It is known that to imperfectly clone a phase-set of states uncorrelatedly is impossible due to the linearity and the hermitian-preservation of quantum operations deterministically  shown by D'Ariano et al. In this paper we address uncorrelated cloning in probabilistic cases.
We show that there exists a linear and hermitian-preserving probabilistic map to imperfectly clone a phase-set of states uncorrelatedly. It is pointed out that the existence of such a map is due to the difference between non-unit-trace output operators and normalized (unit-trace) output operators inherent to probabilistic maps.  We however prove that such a possibility of uncorrelated cloning is completely excluded by the condition of positivity in addition to the linearity and the hermitian-preservation of quantum operations. Our results implicate that the positivity of quantum operation is richer physical meaning than the one which we usually assume {\it a priori} for the necessity of ``probability interpretation.''
\end{abstract}

\maketitle

\section{Introduction}

The time evolution in quantum mechanics has lots of potential for information
processing, on the other hand, it is also known that there exists 
the strict no-go rules in it \cite{bookb, bookc, buhi, scib}, so
analyzing them
leads to a better understanding of quantum mechanics.
In the development of quantum information science, it was mathematically proved that the most generalized time evolution allowed in quantum mechanics is linear and completely positive maps \cite{booki}, which are also called {\it quantum operations}.
The complete positivity is the condition to guarantee the positivity for extended quantum systems, and it automatically includes the hermitian-preservation and the positivity. 
The well-known no-cloning theorem can be regarded as one of the
direct consequences from the linearity of quantum operations \cite{bookb, bookc}, and triggered off the field called ``quantum cloning,'' where one addresses the variety of imperfect cloning from both qualitative and quantitative manner \cite{buhi, scib}. 
In the flow of it, D'Ariano et al. considered {\it uncorrelated cloning}, where all the multiple output states $\rho_{i}$ depend on input states $|\psi\rangle$ and are uncorrelated each other, i.e., $\rho_1 (\psi)\otimes  \rho_2 (\psi)\otimes \cdots \otimes \rho_M (\psi)$, and they showed that it is impossible to uncorrelatedly clone a phase-set of input states deterministically by using the linearity and the hermitian-preservation of quantum operations \cite{dade}. 
A phase-set of states means the set of pure states
\begin{eqnarray}
	\vert \phi\rangle =\sqrt{q}|0\rangle + \sqrt{1-q}e^{i\phi}|1\rangle\label{nea}
\end{eqnarray} 
for continuous range of $\phi$ and a fixed real number $q$ within $0<q<1$ \cite{dade}.

In the present paper, we address the issue of uncorrelated cloning
in non-trace-preserving or {\it probabilistic} cases. 
We show that there exists a probabilistic
 uncorrelated map satisfying the linearity and the hermitian-preservation for a phase-set of input states $\vert \phi\rangle$ for all $\phi$ $(0\leq\phi <2 \pi)$ and both two output states $\rho_1$, $\rho_2$ depend on input states, i.e., $\rho_1 (\phi)\otimes  \rho_2 (\phi)$. We next point out that the existence of such a map is due to
the appearance of probabilities $P(\phi)$ attached to $\rho_{i}(\phi)\ (i=1, 2)$, i.e., $P(\phi)\rho_{i}(\phi)$ inherent to probabilistic maps. 
From this phenomenon,   
we derive the {\it anomalous relation} of probabilistic uncorrelated maps, which is not possessed in deterministic uncorrelated maps. 

Despite such a possibility,
we show that all the maps to uncorreletedly clone $|\phi\rangle$ are not positive maps, 
i.e., $P(\phi)\rho_{i}(\phi)\not\ge 0$. 
In other words, we prove that the possibility of probabilistic uncorrelated cloning 
is excluded by 
the ``positivity'' in addition to the linearity and the hermitian-preservation of quantum operations. Our result means that the positivity plays the essential role to derive the impossibility of probabilistic uncorrelated cloning.

\section{Impossibility of deterministic uncorrelated cloning in linear and hermitian-preserving maps}\label{idu}
\newcommand{\1}{\mbox{1}\hspace{-0.25em}\mbox{l}}
We start with the review of the proof of the impossibility of uncorrelated cloning for a phase-set of states $|\phi\rangle$ proposed by D'Ariano et al. \cite{dade}, which is completely valid for all the deterministic cases. We firstly assume that there exist linear 
hermitian-preserving deterministic (i.e., trace-preserving) maps $\Lambda _{12}, \Lambda _{1}, \Lambda _{2}$, and both two output 
states $\rho_{1}(\phi)$ and $\rho_{2}(\phi)$
of $\Lambda _{1}$ and $\Lambda _{2}$ respectively depend on input states $|\phi\rangle$ and are uncorrelated each other, namely,
\begin{numcases}{}
\vert \phi\rangle \xrightarrow{\Lambda _{12}}\Lambda _{12}(\phi )=
\rho_{1}(\phi )\otimes\rho_{2}(\phi )\label{ned}\\
\vert \phi\rangle \xrightarrow{\Lambda _{1}}\Lambda _{1}(\phi )\equiv\hbox{Tr}_{2}\bigl[\Lambda _{12}(\phi )\bigr]=
\rho_{1}(\phi )\ \ \ \ \ \ \ \ \ \Bigl(\hbox{Tr}\bigl[\rho_{1}(\phi )\bigr]=1\Bigr)\label{neb}\\
\vert \phi\rangle \xrightarrow{\Lambda _{2}}\Lambda _{2}(\phi )\equiv\hbox{Tr}_{1}\bigl[\Lambda _{12}(\phi )\bigr]=
\rho_{2}(\phi )\ \ \ \ \ \ \ \ \ \Bigl(\hbox{Tr}\bigl[\rho_{2}(\phi )\bigr]=1\Bigr).\label{nec}
\end{numcases}
Here a set of input states $|\phi\rangle$ is given as a phase-set of states (\ref{nea}),
and it can be represented with a density matrix
\begin{eqnarray}
\vert \phi\rangle\langle\phi\vert=\left(
\begin{array}{cc}
q & \sqrt{q(1-q)}e^{-i\phi} \\
\sqrt{q(1-q)}e^{i\phi} & 1-q \\
\end{array}
\right)\label{nee}.
\end{eqnarray} 
From the linearity of $\Lambda _{1}, \Lambda _{2}$ and a matrix (\ref{nee}), the $\phi$-dependence
in $\Lambda_{1}(\phi )$ and $\Lambda_{2}(\phi )$ should appear as having polynomials of 
first order of $e^{\pm i\phi}$ somewhere in both matrix-elements 
$\bigl[\Lambda_{1}(\phi )\bigr]_{ij}$ and $\bigl[\Lambda_{2}(\phi )\bigr]_{\mu\nu}$.
Furthermore, from the hermitian-preservation of $\Lambda _{1}, \Lambda _{2}$, 
both $e^{i\phi}$ and $e^{-i\phi}$ should necessarily appear somewhere
in the both matrix-elements.
By noting these facts and the relation
\begin{eqnarray}
\Lambda _{12}(\phi )=\Lambda _{1}(\phi )\otimes\Lambda _{2}(\phi )\label{nei}
\end{eqnarray}
which is easily confirmed from (\ref{ned}), (\ref{neb}), and (\ref{nec}),
one notices that
polynomials of second order of $e^{\pm i\phi}$, i.e., $e^{\pm 2i\phi}$, must 
exist somewhere in matrix-elements $\bigl[\Lambda _{12}(\phi )\bigr]_{ij\mu\nu}$.
This fact however contradicts to the linearity of $\Lambda _{12}$ for 
some continuous range of $\phi$.\footnote{We briefly refer to the conditions assumed in the above proof.
Although, in the paper \cite{dade}, 
only the linearity of maps $\Lambda _{12}, \Lambda _{1}, \Lambda _{2}$ is explicitly assumed
to derive the impossibility of uncorrelated cloning, one have to also assume 
the hermitian-preservation of maps $\Lambda _{12}, \Lambda _{1}, \Lambda _{2}$ as is done above.
In fact, if one only assumes the linearity without the hermitian-preservation,
there exist linear uncorrelated maps $\Lambda _{12}, \Lambda _{1}, \Lambda _{2}$ satisfying
$\Lambda _{12}(\phi )=\Lambda _{1}(\phi )\otimes\Lambda _{2}(\phi )=\left(
\begin{array}{cccc}
1 & 0 & 0 & 0\\
e^{-i\phi} & 0 & 0 & 0\\
e^{i\phi} & 0 & 0 & 0\\
1 & 0 & 0 & 0
\end{array}
\right)
$,
$\Lambda _{1}(\phi )=\left(
\begin{array}{cc}
1 & 0 \\
e^{i\phi} & 0 \\
\end{array}
\right)
$, 
$\Lambda _{2}(\phi )=\left(
\begin{array}{cc}
1 & 0 \\
e^{-i\phi} & 0 \\
\end{array}
\right)
$.
In this example, the $\phi$-dependence is distributed uncorrelatedly
to $\Lambda _{1}(\phi )$ and $\Lambda _{2}(\phi )$, and then all the matrix-elements of 
$\bigl[\Lambda _{12}(\phi )\bigr]_{ij\mu\nu}$,
$\bigl[\Lambda _{1}(\phi )\bigr]_{ij}$,
and $\bigl[\Lambda _{2}(\phi )\bigr]_{\mu\nu}$
are polynomials of first
order of $e^{\pm i\phi}$.
Therefore
there is no contradiction as appeared in the above proof.
In this sense, we can say that uncorrelated cloning for a phase-set of states is possible within
the only condition of linearity.}

Since we have not used the trace-preserving condition in the above proof,
``is this proof also valid for probabilistic cases?''
To answer this question rigorously, we next analyze the character of 
probabilistic uncorrelated maps.

\section{Possibility of probabilistic uncorrelated cloning in linear and hermitian-preserving maps}\label{cexa}

\subsection{Anomalous relation of probabilistic uncorrelated maps}\label{secou}
In probabilistic uncorrelated maps, we have to be careful of the problem 
setting about uncorrelated cloning itself
because of the existence of probabilistic uncorrelated quantum operations such as
\begin{numcases}{}
\vert \psi\rangle \xrightarrow{\Phi _{12}^{P}}\Phi _{12}^{P}(\psi )=
P(\psi) \bigl( |0\rangle\langle 0|\otimes |0\rangle\langle 0|\bigr)\label{neh}\\
\vert \psi\rangle \xrightarrow{\Phi _{1}^{P}}\Phi _{1}^{P}(\psi )\equiv\hbox{Tr}_{2}\bigl[\Phi _{12}^{P}(\psi )\bigr]=
P(\psi) |0\rangle\langle 0|\label{nefx}\\
\vert \psi\rangle \xrightarrow{\Phi _{2}^{P}}\Phi _{2}^{P}(\psi )\equiv\hbox{Tr}_{1}\bigl[\Phi _{12}^{P}(\psi )\bigr]=
P(\psi) |0\rangle\langle 0|\label{negx},
\end{numcases}
where probabilities are given by
$P(\psi)=|\langle0|\psi\rangle|^{2}$, and the index $P$ 
in the quantum operations $\Phi _{12}^{P}, \Phi _{1}^{P}, \Phi _{2}^{P}$ emphasizes the ``probabilistic
maps.'' These quantum operations $\Phi _{12}^{P}, \Phi _{1}^{P}, \Phi _{2}^{P}$ 
are actually realized 
by von Neumann's projective measurement characterized by the measurement operators
$\bigl\{ |0\rangle\langle 0|, |1\rangle\langle 1|\bigr\}$.
After that measurement, one choose the outcome $|0\rangle\langle 0|$ with the probability $P(\psi)$ in 
System 1, and next prepare the same state as $|0\rangle$ for System 2
while the outcome $|1\rangle\langle 1|$ is thrown away. 
In (\ref{nefx}) and (\ref{negx}), the {\it non-unit-trace output operators} 
$\Phi _{1}^{P}(\psi )$, $\Phi _{2}^{P}(\psi )$ are both dependent on input states $|\psi\rangle$ through the probability $P(\psi)$ depending on input states $|\psi\rangle$
although the operators
$|0\rangle$ are independent of input states $|\psi\rangle$
in both Systems 1 and 2.
Such a possibility of ``indirect''  input states-dependence through the probability $P(\psi)$
in 
the {\it non-unit-trace output operators} $\Phi _{1}^{P}(\psi )$, $\Phi _{2}^{P}(\psi )$
have already discussed elsewhere \cite{seis}.

In the present paper, we rather address the possibility of input states-dependence 
in the {\it normalized operators} or
the {\it output states} $\Lambda _{1}^{P}(\phi )/\hbox{Tr}\bigl[\Lambda _{1}^{P}(\phi)\bigr]$, $\Lambda _{2}^{P}(\phi )/\hbox{Tr}\bigl[\Lambda _{2}^{P}(\phi)\bigr]$ in probabilistic uncorrelated maps $\Lambda _{12}^{P}$, 
$\Lambda _{1}^{P}$, $\Lambda _{2}^{P}$.
When both two output states depend on input states,
we here simply call such phenomenon 
{\it probabilistic uncorrelated cloning}.
If one assumed that probabilistic uncorrelated cloning for input states $|\phi\rangle$ by (\ref{nea})
is possible, the maps to realize it should be given by
\begin{numcases}{}
\vert \phi\rangle \xrightarrow{\Lambda _{12}^{P}}\Lambda _{12}^{P}(\phi )=
P(\phi)\bigl[ \rho_{1}(\phi )\otimes\rho_{2}(\phi )\bigr]\label{neh}\\
\vert \phi\rangle \xrightarrow{\Lambda _{1}^{P}}\Lambda _{1}^{P}(\phi )\equiv\hbox{Tr}_{2}\bigl[\Lambda _{12}^{P}(\phi )\bigr]=
P(\phi)\rho_{1}(\phi )\label{nef}\\
\vert \phi\rangle \xrightarrow{\Lambda _{2}^{P}}\Lambda _{2}^{P}(\phi )\equiv\hbox{Tr}_{1}\bigl[\Lambda _{12}^{P}(\phi )\bigr]=
P(\phi)\rho_{2}(\phi ),\label{neg}
\end{numcases}
where probabilities $P(\phi)$ are defined as 
\begin{eqnarray}
P(\phi) \equiv \hbox{Tr}\bigl[\Lambda _{1}^{P}(\phi)\bigr]
=\hbox{Tr}\bigl[\Lambda _{2}^{P}(\phi)\bigr]
=\hbox{Tr}\bigl[\Lambda _{12}^{P}(\phi)\bigr],  \label{neha}
\end{eqnarray}
and both two output states $\rho_{1}(\phi )$ and $\rho_{2}(\phi )$ depend on $\phi$.

From the right-hand sides of (\ref{neh}), (\ref{nef}), (\ref{neg}), one can find that the relation
\begin{eqnarray}
\Lambda _{12}^{P}(\phi )=\frac{\Lambda _{1}^{P}(\phi )\otimes\Lambda _{2}^{P}(\phi )}{P(\phi )}\label{nej}
\end{eqnarray}
holds. Without the loss of generality, we only consider the case $P(\phi )\ne 0$.\footnote{When $P(\phi )=0$, there are no outputs in any systems. So we need not consider such a case.}
In the sense that a probability $P(\phi)$ appears in the denominator of the right-hand side of (\ref{nej}),
which never appears in the relation corresponding to deterministic cases (\ref{nei}),
we can say that probabilistic uncorrelated maps generically have an {\it anomalous relation}.
If one imposes the linearity on probabilistic maps
$\Lambda _{12}^{P}$, $\Lambda _{1}^{P}$, $\Lambda _{2}^{P}$, 
a probability $P(\phi)$ becomes a polynomial of first order of $e^{\pm i\phi}$ by 
its definition (\ref{neha}).
Therefore it seems that the anomalous relation (\ref{nej}) might enable
all the matrix-elements $\bigl[\Lambda _{12}^{P}(\phi )\bigr]_{ij\mu\nu}$
to be polynomials of first order of $e^{\pm i\phi}$ because
polynomials of second order of $e^{\pm i\phi}$ appeared in
matrix-elements in the numerator of the right-hand side of (\ref{nej}) 
could reduce to polynomials of first order of $e^{\pm i\phi}$
by a probability $P(\phi)$ in the denominator which is a polynomial of 
first order of $e^{\pm i\phi}$.
From this consideration, it can be said that the anomalous relation (\ref{nej})
makes applying
the proof of the impossibility of deterministic uncorrelated cloning 
for probabilistic cases unreliable.
And in fact, we can explicitly propose a 
counterexample for such a applicability to the probabilistic cases 
although that applicability is briefly mentioned in \cite{dade}.

\subsection{Counterexample}\label{cou}
For a phase-set of states $|\phi\rangle$, we suggest the linear and hermitian-preserving uncorrelated maps
\begin{numcases}{}
\vert \phi\rangle \xrightarrow{\tilde{\Lambda} _{12}^{P}}\tilde{\Lambda} _{12}^{P}(\phi )=\frac{\tilde{\Lambda} _{1}^{P}(\phi )\otimes\tilde{\Lambda} _{2}^{P}(\phi )}{\tilde{P}(\phi )}\ \ \ \ \ \ \ \ \ \ \ \ \ \ \ (\tilde{P}(\phi)\ne 0)\label{nem}\\
\ \ \ \ \ \ \ \ \ \ \ \ \ \ \ \ \ \ \ \ =\left(
\begin{array}{cccc}
\frac{1}{16}\scriptstyle{(e^{i\phi}+1)(e^{-i\phi}+1)}&\frac{1}{2}(e^{-i\phi}+1)&\frac{1}{2}(e^{-i\phi}+1)&4e^{-i\phi}\\
\frac{1}{2}(e^{i\phi}+1)&\frac{1}{16}\scriptstyle{(e^{i\phi}+1)(e^{-i\phi}+1)}&4&\frac{1}{2}(e^{-i\phi}+1)\\
\frac{1}{2}(e^{i\phi}+1)&4&\frac{1}{16}\scriptstyle{(e^{i\phi}+1)(e^{-i\phi}+1)}&\frac{1}{2}(e^{-i\phi}+1)\\
4e^{i\phi}&\frac{1}{2}(e^{i\phi}+1)&\frac{1}{2}(e^{i\phi}+1)&\frac{1}{16}\scriptstyle{(e^{i\phi}+1)(e^{-i\phi}+1)}  
\end{array} 
\right)  \nonumber\\
\vert \phi\rangle \xrightarrow{\tilde{\Lambda} _{1}^{P}}\tilde{\Lambda} _{1}^{P}(\phi )=
\left(
\begin{array}{cc}
\displaystyle\frac{1}{8} (e^{i\phi}+1)(e^{-i\phi}+1)  &  e^{-i\phi}+1    \\
e^{i\phi}+1 &  \displaystyle\frac{1}{8} (e^{i\phi}+1)(e^{-i\phi}+1)         
\end{array} 
\right) \label{nek}\\
\vert \phi\rangle \xrightarrow{\tilde{\Lambda} _{2}^{P}}\tilde{\Lambda} _{2}^{P}(\phi )=
\left(
\begin{array}{cc}
\displaystyle\frac{1}{8} (e^{i\phi}+1)(e^{-i\phi}+1)  &  e^{-i\phi}+1    \\
e^{i\phi}+1 &  \displaystyle\frac{1}{8} (e^{i\phi}+1)(e^{-i\phi}+1)         
\end{array} 
\right), \label{nel}
\end{numcases}
where a probability $\tilde{P}(\phi)$ to realize output states is given by 
\begin{eqnarray}
\tilde{P}(\phi) &\equiv &\hbox{Tr}\bigl[\tilde{\Lambda} _{1}^{P}(\phi)\bigr]
=\hbox{Tr}\bigl[\tilde{\Lambda} _{2}^{P}(\phi)\bigr]
=\hbox{Tr}\bigl[\tilde{\Lambda} _{12}^{P}(\phi)\bigr]  \nonumber\\
&=&\displaystyle\frac{1}{4} (e^{i\phi}+1)(e^{-i\phi}+1). \label{nen}
\end{eqnarray}
(There are no outputs when $\tilde{P}(\phi)=0\ (\phi=\pi)$).
One should note that all the matrix-elements of the right-hand sides of (\ref{nem}), (\ref{nek}), (\ref{nel}) 
are polynomials of first order of $e^{\pm i\phi}$, therefore the linearity of $\tilde{\Lambda} _{12}^{P}$,
$\tilde{\Lambda} _{1}^{P}$, $\tilde{\Lambda} _{2}^{P}$
is satisfied. And the hermitian-preservation of $\tilde{\Lambda} _{12}^{P}$,
$\tilde{\Lambda} _{1}^{P}$, $\tilde{\Lambda} _{2}^{P}$
is also satisfied because, among the matrix-elements, the relations 
$\bigl[\tilde{\Lambda}_{1}(\phi )\bigr]_{ji}=\bigl[\tilde{\Lambda}_{1}(\phi )\bigr]^{\ast}_{ij}$, $\bigl[\tilde{\Lambda}_{2}(\phi )\bigr]_{nm}=\bigl[\tilde{\Lambda}_{2}(\phi )\bigr]^{\ast}_{mn}$, $\bigl[\tilde{\Lambda}_{12}(\phi )\bigr]_{jinm}=\bigl[\tilde{\Lambda}_{12}(\phi )\bigr]^{\ast}_{ijmn}$ hold for all $i, j, m, n$.
Since, from the right-hand sides of (\ref{nek}), (\ref{nel}), the output states
$\tilde{\rho} _{1}(\phi )\equiv\tilde{\Lambda} _{1}(\phi )/\tilde{P}(\phi)$, 
$\tilde{\rho} _{2}(\phi )\equiv\tilde{\Lambda} _{2}(\phi )/\tilde{P}(\phi)$ are given by
\begin{eqnarray}
\tilde{\rho} _{1}(\phi )=\tilde{\rho} _{2}(\phi )=
\left(
\begin{array}{cc}
\frac{1}{2} &  4(e^{i\phi}+1)^{-1}    \\
4(e^{-i\phi}+1)^{-1} &  \frac{1}{2}         
\end{array} 
\right),  \label{neo}
\end{eqnarray}
both the output states $\tilde{\rho} _{1}(\phi )$, $\tilde{\rho} _{2}(\phi )$
depend on $\phi$, and they are uncorrelated each other from (\ref{nem}).
Therefore these maps $\tilde{\Lambda} _{12}^{P}$,
$\tilde{\Lambda} _{1}^{P}$, $\tilde{\Lambda} _{2}^{P}$ indicates that {\it probabilistic uncorrelated cloning is possible
on the condition of the linearity and the hermitian-preservation,
and then one can recognize the maps} (\ref{nem}), (\ref{nek}), (\ref{nel}) {\it as the counterexample
to applying the proof of the impossibility of deterministic uncorrelated cloning 
for probabilistic cases}.

\subsection{General analysis of probabilistic uncorrelated maps}
To analyze the generality about the above counterexample, we express the relation (\ref{nej}) in probabilistic uncorrelated maps by matrix-elements like
\begin{eqnarray}
P(\phi) \cdot\bigl[\Lambda _{12}^{P}(\phi )\bigr]_{ij\mu\nu}=\bigl[\Lambda _{1}^{P}(\phi )\bigr]_{ij}\cdot\bigl[\Lambda _{2}^{P}(\phi )\bigr]_{\mu\nu}.\label{dirhh}
\end{eqnarray}
Here we assume that $\Lambda _{12}^{P}$, $\Lambda _{1}^{P}$, and $\Lambda _{2}^{P}$ are linear and hermitian-preserving probabilistic maps.
If we consider that a
probability $P(\phi)$ are some fixed (i.e., independent of $\phi$) one, the proof of the 
impossibility of deterministic uncorrelated cloning 
can be also applicable to such restricted probabilistic cases
because a fixed probability $P$ cannot reduce 
polynomials of second order of $e^{\pm i\phi}$ appeared in
matrix-elements in the numerator of the right-hand side of (\ref{nej}) 
to polynomials of first order of $e^{\pm i\phi}$,
which is also acknowledged in the above relation (\ref{dirhh}).

The assumption of the linearity and the hermitian-preservation of $\Lambda _{12}^{P}$, 
$\Lambda _{1}^{P}$, $\Lambda _{2}^{P}$, and the definition (\ref{neha}) in general 
make
a probability $P(\phi)$ to be a real polynomial of first order of $e^{\pm i\phi}$
as seen in the counterexample in Sec. \ref{cou}. Therefore 
a probability $P(\phi)$ in (\ref{dirhh}) can be generically expressed as follows.
\begin{eqnarray}
P(\phi)=\alpha e^{i\phi}+ \alpha^{\ast}e^{-i\phi}+R \ \ \ \ \ \ \ \ (\alpha\in \mathbb{C},\ R\in \mathbb{R}),  \label{diry}
\end{eqnarray}
where $\alpha$ and $R$ are constants independent of $\phi$.v
From the above discussion, since there is no possibility of uncorrelated cloning
when a probability $P(\phi)$ is independent of $\phi$,
we only consider the probabilistic uncorrelated maps whose probabilities
$P(\phi)$ depend on $\phi$, 
namely, a coefficient $\alpha$ in (\ref{diry}) is not 0. Such a probability $P(\phi)$ in (\ref{diry})
can be represented by
the {\it decomposition form} in (\ref{dirp}) as follows:
\begin{eqnarray}
P(\phi)=\alpha (e^{i\phi}+p_{0})(p_{1}e^{-i\phi}+1)\ \ \ \ (\alpha\ne 0), \label{neq} 
\end{eqnarray}
where $p_{0}$ and $p_{1}$ are non-zero
constants independent of $\phi$ which are uniquely determined up to a permutation (see Appendix) to hold the equation (\ref{diry}). Similarly,
the linearity of $\Lambda _{12}^{P}$, $\Lambda _{1}^{P}$, and $\Lambda _{2}^{P}$
also allows us to represent all the matrix-elements 
$\bigl[\Lambda _{12}^{P}(\phi )\bigr]_{ij\mu\nu}$,
$\bigl[\Lambda _{1}^{P}(\phi )\bigr]_{ij}$,
$\bigl[\Lambda _{2}^{P}(\phi )\bigr]_{\mu\nu}$
by any one of  {\it decomposition forms}  
(\ref{dirp}), (\ref{dirs}), (\ref{dirt}).

As the preparation for making the analysis of the relation (\ref{dirhh}) clear,
we multiply a probability $P(\phi)$ and all the matrix-elements 
$\bigl[\Lambda _{12}^{P}(\phi )\bigr]_{ij\mu\nu}$,
$\bigl[\Lambda _{1}^{P}(\phi )\bigr]_{ij}$,
$\bigl[\Lambda _{2}^{P}(\phi )\bigr]_{\mu\nu}$ in decomposition forms
by $e^{i\phi}$,
i.e., $e^{i\phi}P(\phi)$,
$e^{i\phi}\bigl[\Lambda _{12}^{P}(\phi )\bigr]_{ij\mu\nu}$,
$e^{i\phi}\bigl[\Lambda _{1}^{P}(\phi )\bigr]_{ij}$,
$e^{i\phi}\bigl[\Lambda _{2}^{P}(\phi )\bigr]_{\mu\nu}$,
all become factored-polynomials of $e^{ i\phi}$ of an order 0, 1, or 2. 
For instance, when an order is 2,
such a
factored-polynomial is given by
$\ast(e^{i\phi}+\ast\ast)(e^{i\phi}+\ast\ast\ast)$,
where
each of symbols ``$\ast$,'' ``$\ast\ast$,'' ``$\ast\ast\ast$'' indicates some fixed complex number, and each of $(e^{i\phi}+\ast\ast)$, $(e^{i\phi}+\ast\ast\ast)$ is simply called a {\it factor} here. 
Then, for instance, a factor $(e^{i\phi}+\ast\ast)$ is the same as the other factor
if and only if a number ``$\ast\ast$'' coincides with the other's one.
As discussed below, 
the form of factored-polynomials
contributes to make the discussion easy to handle and visible.

By the above preparation and multiplying (\ref{dirhh}) by $e^{2i\phi}$, we have a factored-polynomial
of $e^{ i\phi}$ of an order 2, 3, or 4 as follows:
\begin{eqnarray}
e^{i\phi}P(\phi)\cdot  e^{i\phi}\bigl[\Lambda _{12}^{P}(\phi )\bigr]_{ij\mu\nu}=e^{i\phi}\bigl[\Lambda _{1}^{P}(\phi )\bigr]_{ij}\cdot e^{i\phi}\bigl[\Lambda _{2}^{P}(\phi )\bigr]_{\mu\nu}
=\displaystyle\prod_{k=0}^{L_{ij\mu\nu}} u_{ij\mu\nu}(e^{ i\phi}+\omega_{ij\mu\nu}^{k}),\label{ner}\\ 
\ \ \ \ \ \ \ (1\leq L_{ij\mu\nu}\leq 3, \ \ L_{ij\mu\nu}\in \mathbb{Z})\nonumber
\end{eqnarray}
where $u_{ij\mu\nu}$ is some fixed (i.e., independent of $\phi$) coefficient,
and $\omega_{ij\mu\nu}^{k}$ are also fixed values
which are uniquely determined up to a permutation
by the uniqueness of polynomial factorization in algebra \cite{bookj} (see also Appendix).

Now, since a probability $P(\phi)$ in the decomposition form is given by (\ref{neq}), two out of $\omega_{ij\mu\nu}^{k}$ in (\ref{ner}) 
must be $p_{0}$ and $p_{1}$. This fact allows us to choose 
$\omega_{ij\mu\nu}^{0}$ and $\omega_{ij\mu\nu}^{1}$ as
\begin{numcases}{}
		\omega_{ij\mu\nu}^{0}=p_{0}\ \ \ \ \ \ \ \ \ (\forall\ i, j,\mu, \nu) \label{dirjj}\\
		\omega_{ij\mu\nu}^{1}=p_{1}\ \ \ \ \ \ \ \ \ (\forall\ i, j,\mu, \nu) \label{dirkk}
\end{numcases}
without the loss of generality.
The value of integer $L_{ij\mu\nu}$ is determined by which {\it decomposition form} each
$\bigl[\Lambda _{12}^{P}(\phi )\bigr]_{ij\mu\nu}$ takes: When 
$\bigl[\Lambda _{12}^{P}(\phi )\bigr]_{ij\mu\nu}$ takes the form (\ref{dirt}),
namely, 
\begin{eqnarray}
\bigl[\Lambda _{12}^{P}(\phi )\bigr]_{ij\mu\nu}=\tilde{b}e^{-i\phi} \ \ \ \ \ \ \ \ \ \ (\tilde{b}\in \mathbb{C}),
\end{eqnarray}
$L_{ij\mu\nu}=1$. When 
$\bigl[\Lambda _{12}^{P}(\phi )\bigr]_{ij\mu\nu}$ takes the form (\ref{dirs}),
namely, 
\begin{eqnarray}
\bigl[\Lambda _{12}^{P}(\phi )\bigr]_{ij\mu\nu}=\tilde{c}(\tilde{w}e^{-i\phi}+1) \ \ \ \ \ \ \ \ \ \ (\tilde{w}, \tilde{c}\in \mathbb{C},\  \tilde{c}\ne 0),
\end{eqnarray}
$L_{ij\mu\nu}=2$. When 
$\bigl[\Lambda _{12}^{P}(\phi )\bigr]_{ij\mu\nu}$ takes the form (\ref{dirp}),
namely, 
\begin{eqnarray}
\bigl[\Lambda _{12}^{P}(\phi )\bigr]_{ij\mu\nu}=\tilde{a} (e^{i\phi}+\tilde{w}_{0})(\tilde{w}_{1}e^{-i\phi}+1)\ \ \ \ \ \ \ \ \ \ \ \ (\tilde{w}_{0}, \tilde{w}_{1}, \tilde{a}\in \mathbb{C},\  \tilde{a}\ne 0),
\end{eqnarray}
$L_{ij\mu\nu}=3$.

From the above discussion, two factors $(e^{i\phi}+p_{0})$ and 
$(e^{i\phi}+p_{1})$ originated from $e^{i\phi}P(\phi)$ 
must be inlaid as factors of factored-polynomials
$e^{i\phi}\bigl[\Lambda _{1}^{P}(\phi )\bigr]_{ij}$,
$e^{i\phi}\bigl[\Lambda _{2}^{P}(\phi )\bigr]_{\mu\nu}$ by the middle part of the 
relation (\ref{ner})
and the uniqueness of polynomial factorization.\footnote{Note that this classical renowned fact in algebra is valid for any orders of polynomials \cite{bookj}.}
Actually,
one can divide in three cases as below how to inlay two factors
$(e^{i\phi}+p_{0})$,
$(e^{i\phi}+p_{1})$ into $e^{i\phi}\bigl[\Lambda _{2}^{P}(\phi )\bigr]_{\mu\nu}$.

{\center \subsubsection*{Case 1}}

Suppose that {\it some} non-zero factored polynomials $e^{i\phi}\bigl[\Lambda _{2}^{P}(\phi )\bigr]_{\mu\nu}$
contain neither of two factors $(e^{i\phi}+p_{0})$, $(e^{i\phi}+p_{1})$.
In this case, {\it all} the factored-polynomials
$e^{i\phi}\bigl[\Lambda _{1}^{P}(\phi )\bigr]_{ij}$ must contain both two factors $(e^{i\phi}+p_{0})$, $(e^{i\phi}+p_{1})$ to hold the relation (\ref{ner}), namely,
\begin{eqnarray}
e^{i\phi}\bigl[\Lambda _{1}^{P}(\phi )\bigr]_{ij}=s_{ij}(e^{i\phi}+p_{0})(e^{i\phi}+p_{1})\ \ \ \ \ \ \ \ (\forall\ i, j),\label{nes}
\end{eqnarray}
where $s_{ij}$ is some fixed (i.e., independent of $\phi$) coefficient.
The above equation (\ref{nes}) is equivalent to
\begin{eqnarray}
\Lambda _{1}^{P}(\phi)= P(\phi)\Gamma_{1},  \label{dirmm}
\end{eqnarray} 
where $\Gamma_{1}$ is a fixed (i.e., independent of $\phi$) unit-trace operator, and
its elements are given by $\bigl[\Gamma_{1}\bigr]_{ij}=s_{ij}/\alpha$ from (\ref{neq}).
Since the unit-trace operator $\Gamma_{1}=\Lambda _{1}^{P}(\phi)/ P(\phi)$
is independent of $\phi$, it is impossible to uncorrelatedly clone input states 
$|\phi\rangle$ in Case 1.

If one wants to make the existence of Case 1 explicit, for instance, one can come up with the linear and
hermitian-preserving
maps $\Lambda _{12}^{\prime P}$, $\Lambda _{1}^{\prime P}$, $\Lambda _{2}^{\prime P}$ to realize this case as below.
\begin{numcases}{}
\vert \phi\rangle \xrightarrow{\Lambda _{12}^{\prime P}}\Lambda_{12}^{\prime P}(\phi )=\frac{\Lambda _{1}^{\prime P}(\phi )\otimes\Lambda _{2}^{\prime P}(\phi )}{P^{\prime}(\phi )}\label{net}\\
\ \ \ \ \ \ \ \ \ \ \ \ \ \ \ \ \ \ \ \ =\left(
\begin{array}{cccc}
\displaystyle\frac{1}{2} P^{\prime}(\phi)-\displaystyle\frac{1}{8}&0&0&0\\
0&\displaystyle\frac{1}{8}&0&0\\
0&0&\displaystyle\frac{1}{2} P^{\prime}(\phi)-\displaystyle\frac{1}{8}&0\\
0&0&0&\displaystyle\frac{1}{8} 
\end{array} 
\right)  \nonumber\\
\vert \phi\rangle \xrightarrow{\Lambda _{1}^{\prime P}}\Lambda _{1}^{\prime P}(\phi )=
\left(
\begin{array}{cc}
\displaystyle\frac{1}{2} P^{\prime}(\phi)  &  0    \\
0&  \displaystyle\frac{1}{2} P^{\prime}(\phi)        
\end{array} 
\right) \label{neu}\\
\vert \phi\rangle \xrightarrow{\Lambda _{2}^{\prime P}}\Lambda _{2}^{\prime P}(\phi )=
\left(
\begin{array}{cc}
P^{\prime}(\phi)-\displaystyle\frac{1}{4} &  0   \\
0 &  \displaystyle\frac{1}{4}     
\end{array} 
\right), \label{nev}
\end{numcases}
where the probability $P^{\prime}(\phi)$ is given by 
\begin{eqnarray}
P^{\prime}(\phi) &\equiv &\hbox{Tr}\bigl[\Lambda_{1}^{\prime P}(\phi)\bigr]
=\hbox{Tr}\bigl[\Lambda _{2}^{\prime P}(\phi)\bigr]
=\hbox{Tr}\bigl[\Lambda _{12}^{\prime P}(\phi)\bigr]  \nonumber\\
&=&\displaystyle\frac{1}{8} (e^{i\phi}+2+\sqrt{3})\bigl[(2-\sqrt{3})e^{-i\phi}+1\bigr]. \label{new}
\end{eqnarray}
Since the polynomial $e^{i\phi}(\phi)\bigl[\Lambda _{2}^{\prime P}(\phi )\bigr]_{11}$ can be written as a factored-polynomial $1/8(e^{i\phi}+1)(e^{-i\phi}+1)$, one can indeed make sure that there are neither of two factors
$(e^{i\phi}+2+\sqrt{3})$, $(e^{i\phi}+2-\sqrt{3})$ of $e^{i\phi}P^{\prime}(\phi)$
in $e^{i\phi}\bigl[\Lambda _{2}^{\prime P}(\phi )\bigr]_{11}$. 
Therefore the above maps are considered as Case 1. And then the non-zero factored-polynomials
$e^{i\phi}\bigl[\Lambda _{1}^{\prime P}(\phi )\bigr]_{11}$,
$e^{i\phi}\bigl[\Lambda _{1}^{\prime P}(\phi )\bigr]_{22}$
have both two factors
$(e^{i\phi}+2+\sqrt{3})$, $(e^{i\phi}+2-\sqrt{3})$ of $e^{i\phi}P^{\prime}(\phi)$,
so the unit-trace operator $\Lambda _{1}^{\prime P}(\phi )/P^{\prime}(\phi)=\1/2$
is independent of $\phi$.
\\
\\

{\center \subsubsection*{Case 2}}

Suppose that {\it some} non-zero factored-polynomials $e^{i\phi}\bigl[\Lambda _{2}^{P}(\phi )\bigr]_{\mu\nu}$
contain either one of two factors $(e^{i\phi}+p_{0})$, $(e^{i\phi}+p_{1})$.
In this case, {\it all} the factored-polynomials
$e^{i\phi}\bigl[\Lambda _{1}^{P}(\phi )\bigr]_{ij}$ must contain the other factor
which is not contained in $e^{i\phi}\bigl[\Lambda _{2}^{P}(\phi )\bigr]_{\mu\nu}$ as a factor to hold the equation (\ref{ner}). Without the loss of generality, one may consider that 
$e^{i\phi}\bigl[\Lambda _{2}^{P}(\phi )\bigr]_{\mu\nu}$ contains a factor $(e^{i\phi}+p_{1})$, and then all the factored-polynomials $e^{i\phi}\bigl[\Lambda _{1}^{P}(\phi )\bigr]_{ij}$ can be written as
\begin{eqnarray}
e^{i\phi}\bigl[\Lambda _{1}^{P}(\phi )\bigr]_{ij}
=s_{ij}\displaystyle\prod_{l=0}^{M_{ij}} (e^{i\phi}+f_{ij}^{l})\ \ \ \ \ (0\leq M_{ij}\leq 1, \ \ M_{ij}\in \mathbb{Z})\label{dirnn}
\end{eqnarray}
where $f_{ij}^{0}=p_{0}$, and $f_{ij}^{1}$ is some fixed complex number 
which appears only when $M_{ij}=1$.
The value of integer $M_{ij}$ is determined by which {\it decomposition form} each
$\bigl[\Lambda _{1}^{P}(\phi )\bigr]_{ij}$ takes: When 
$\bigl[\Lambda _{1}^{P}(\phi )\bigr]_{ij}$ takes the form (\ref{dirs}),
$M_{ij}=0$, and when 
$\bigl[\Lambda _{1}^{P}(\phi )\bigr]_{ij}$ takes the form (\ref{dirp}),
$M_{ij}=1$.

If $M_{ij}= 1$ and $f_{ij}^{1}=p_{1}$, i.e., $\bigl[\Lambda _{1}^{P}(\phi )\bigr]_{ij}=s_{ij}(e^{i\phi}+p_{0})(p_{1}e^{i\phi}+1)$, 
a formula $\bigl[\Lambda _{1}^{P}(\phi )\bigr]_{ij}/P(\phi)$ is independent of $\phi$
because both two factors $(e^{i\phi}+p_{0})$, $(e^{i\phi}+p_{1})$ 
appeared in the numerator $\bigl[\Lambda _{1}^{P}(\phi )\bigr]_{ij}$ 
are canceled by the denominator $P(\phi)$.
Therefore, if there exists at least one factored formula $e^{i\phi}\bigl[\Lambda _{1}^{P}(\phi )\bigr]_{ij}$ which is not such a case, namely, when 
$M_{ij}=0$ or $f_{ij}^{1}\ne p_{1}$,
a unit-trace operator $\Lambda _{1}^{P}(\phi )/P(\phi)$ depends on $\phi$.
By this reason
one should also note that a unit-trace operator $\Lambda _{2}^{P}(\phi )/P(\phi)$ 
depends on $\phi$ in Case 2.
So, when $M_{ij}=0$ or $f_{ij}^{1}\ne p_{1}$ for some $i, j$,
this leads to {\it uncorrelated cloning}, which is certainly realized by the 
counterexample in Sec. \ref{cou}.

By the way, while the maps $\tilde{\Lambda} _{12}^{P}$, $\tilde{\Lambda} _{1}^{P}$,
$\tilde{\Lambda} _{2}^{P}$ in (\ref{nem}), (\ref{nek}), and (\ref{nel})
of the counterexample in Sec. \ref{cou}
are linear and hermitian-preserving, one can easily check that they are not 
positive maps.
In the next section, we will generically address this issue and
show that all the maps $\Lambda _{1}^{P}$
in the possible counterexamples are not positive by the contraposition:
{\it The additional assumption ``positivity''
to the linearity and hermitian-preservation of maps $\Lambda _{12}^{P}$, $\Lambda _{1}^{P}$,
$\Lambda _{2}^{P}$ 
necessarily implies 
$M_{ij}= 1$ and $f_{ij}^{1}=p_{1}$ for all $i, j$ in} (\ref{dirnn}).
{\it Thus a possibility of uncorrelated cloning induced in Case 2 is excluded.}
\\
\\

{\center \subsubsection*{Case 3}}

This is a remaining case except for Cases 1 and 2:
all the non-zero $e^{i\phi}\bigl[\Lambda _{2}^{P}(\phi )\bigr]_{\mu\nu}$
contain both two factors $(e^{i\phi}+p_{0})$, $(e^{i\phi}+p_{1})$.
One can immediately recognize that the same discussion in Case 1 holds in this case,
thus one eventually has the following relation
\begin{eqnarray}
\Lambda _{2}^{P}(\phi)= P(\phi)\Gamma_{2},  \label{nex}
\end{eqnarray} 
where $\Gamma_{2}$ is a fixed unit-trace operator, instead of (\ref{dirmm}) in this case.
Therefore it is impossible to uncorrelatedly clone input states 
$|\phi\rangle$ in Case 3 as well as Case 1.

One can notice that the example in Case 1 also realizes Case 3 by interchanging System 1 with System 2 for instance.
\\
\\

\section{Impossibility of probabilistic uncorrelated cloning in linear, hermitian-preserving, and positive maps}\label{pro}

Within the conditions of linearity and hermitian-preservation, 
there are maps 
to uncorrelatedly clone a phase-set of states $|\phi\rangle$ from the discussion
of Section \ref{cexa}
and we recognize that all the possibility of uncorrelated cloning comes from Case 2
with $M_{ij}=0$ or $f_{ij}^{1}\ne p_{1}$ for some $i, j$.
In this section, we consider that possibility with the additional condition ``positivity'' to
linear and hermitian-preserving 
maps. By using this additional condition, we derive
$M_{ij}= 1$ and $f_{ij}^{1}=p_{1}$ for all $i, j$ in (\ref{dirnn}), 
i.e., $\bigl[\Lambda _{1}^{P}(\phi )\bigr]_{ij}=(s_{ij}/\alpha)\cdot P(\phi)\ (\forall i,j)$.
Therefore the possibility of uncorrelated cloning occurred in
Case 2 is excluded in linear, hermitian-preserving, and ``positive'' maps.

\subsection{Positivity of $P(\phi)$}\label{prpo}

From the linearity and the hermitian-preservation of maps 
$\Lambda _{12}^{P}$, $\Lambda _{1}^{P}$, $\Lambda _{2}^{P}$,
we obtained a probability given by (\ref{diry}),
and here we firstly rewrite it as
\begin{eqnarray}
P(\phi )= 2\vert \alpha \vert \cos(\phi -\theta ) +R,    \label{diraa}
\end{eqnarray}
where $\alpha\equiv \vert \alpha \vert e^{-i\theta}\ (\vert \alpha \vert \ne 0)$. 
If we additionally assume 
that $\Lambda _{12}^{P}$, $\Lambda _{1}^{P}$, $\Lambda _{2}^{P}$ are positive,
a probability $P(\phi )$ also becomes 
positive for all $\phi$ because $P(\phi )$ is defined as (\ref{neha}).
It is easily confirmed that the inequality
\begin{eqnarray}
\displaystyle\frac{R}{\vert \alpha \vert}\ge 2 \label{dirbb}
\end{eqnarray}
is a necessary and sufficient condition 
for the positivity of $P(\phi )$ in (\ref{diraa}).
By introducing a parameter $r\ (>0)$, the left-hand side of (\ref{dirbb}) can be written as
\begin{eqnarray}
\displaystyle\frac{R}{\vert \alpha \vert}=r+\displaystyle\frac{1}{r}\ \ \ \ \ \ \ (r>0) \label{dircc}
\end{eqnarray}
because a function $r+1/r\ (r>0)$ takes all values larger than 2 uniquely.
Therefore a probability $P(\phi)$ on which the positivity is imposed becomes
\begin{eqnarray}
P(\phi)=\vert \alpha \vert e^{-i\theta}(e^{i\phi}+r e^{i\theta})(r^{-1}e^{i\theta}e^{-i\phi}+1)   \label{dirdd}
\end{eqnarray}
in the decomposition form by using a parameter $r$ satisfying (\ref{dircc}).
The decomposition form of the right-hand side in (\ref{dirdd}) gives the values of constants $p_{0}$, $p_{1}$
in (\ref{neq}) explicitly:
\begin{numcases}{}
		p_{0}=r e^{i\theta} \label{ney}\\
		p_{1}=r^{-1}e^{i\theta}.\label{nez}
\end{numcases}

\subsection{Positivity of $\Lambda _{1}^{P}$}

From (\ref{dirnn}) and (\ref{ney}), one can recognize that
all the elements of $\bigl[\Lambda _{1}^{P}(\phi )\bigr]_{ij}$ in Case 2 come from
two candidates:
\begin{numcases}{\bigl[\Lambda _{1}^{P}(\phi )\bigr]_{ij}=}
		s_{ij}\ (r e^{i\theta}e^{-i\phi}+1)\ \ \ \ \ \ \ \ \ \ \ \ \ \ \ \ \ \ \ \ (when\ M_{ij}=0 ) \label{dirpp}\\
		s_{ij}\ (e^{i\phi}+r e^{i\theta})(f_{ij}e^{-i\phi}+1)\ \ \ \ \ \ \ \ \ (when\ M_{ij}=1). \label{dirqq}
\end{numcases}
Here we define $f_{ij}\equiv f_{ij}^{1}$.
If one impose the positivity on $\Lambda _{1}^{P}$, a $2\times 2$-matrix constructed by 
elements $ii, ij, ji, jj$
\begin{eqnarray}
\left(
\begin{array}{cc}
\bigl[\Lambda _{1}^{P}(\phi )\bigr]_{ii} &  \bigl[\Lambda _{1}^{P}(\phi )\bigr]_{ji}    \\
\bigl[\Lambda _{1}^{P}(\phi )\bigr]_{ij} &  \bigl[\Lambda _{1}^{P}(\phi )\bigr]_{jj}        
\end{array} 
\right)  \label{dirww}
\end{eqnarray}
are also positive. (As shown below, to derive the impossibility of uncorrelated cloning,
it is sufficient to consider the positivity of $2\times 2$-matrices for all $i$, $j$
given in (\ref{dirww})).

The positivity on an above $2\times 2$-matrix leads to
the positivity of all the diagonal elements 
$\bigl[\Lambda _{1}^{P}(\phi )\bigr]_{ii}\ (\forall i)$, so they all have to be in the form (\ref{dirqq}). And then
the calculation imposing the positivity on (\ref{dirqq}) is the same as one in the case
of $P(\phi)$ done in Sec. \ref{prpo}, thus one immediately obtains
$f_{ii}=r^{-1}e^{i\theta}\ (\forall i)$. Therefore
all the diagonal elements become
\begin{eqnarray}
\bigl[\Lambda _{1}^{P}(\phi )\bigr]_{ii}=a_{i}P(\phi )\ \ \ \ \ \ \ \ (a_{i}\geq 0,\ \forall i), \label{dirtt}
\end{eqnarray}
where $a_{i}$ is some positive number.

We next consider non-zero off-diagonal elements $\bigl[\Lambda _{1}^{P}(\phi )\bigr]_{ij}\ (i\ne j, s_{ij}\ne0)$
with the condition $a_{i}a_{j}\ne 0$ in (\ref{dirww}) because a
matrix (\ref{dirww}) is trivially not positive if non-zero off-diagonal elements
exist with $a_{i}a_{j}= 0$. 
If such an off-diagonal element is written in the form (\ref{dirpp}),
the hermiticity of a matrix (\ref{dirww}) implies
\begin{eqnarray}
\bigl[\Lambda _{1}^{P}(\phi )\bigr]_{ji}=\bigl[\Lambda _{1}^{P}(\phi )\bigr]_{ij}^{\ast}=s_{ij}^{\ast}r e^{-i\theta}(e^{i\phi}+r ^{-1}e^{i\theta})\ \ \ (i\ne j,\ s_{ij}\ne0).  \label{diruu}
\end{eqnarray}
This decomposition form $\bigl[\Lambda _{1}^{P}(\phi )\bigr]_{ji}$ is
realized in ($j, i$)-element of the form (\ref{dirqq}) 
when $f_{ji}=0$ and $r=1$.
Thus the diagonal and off-diagonal elements in this case become as follows:
\begin{numcases}{}
\bigl[\Lambda _{1}^{P}(\phi )\bigr]_{ii}=a_{i}\vert \alpha \vert e^{-i\theta}(e^{i\phi}+e^{i\theta})(e^{i\theta}e^{-i\phi}+1) \label{neaa}\\
\bigl[\Lambda _{1}^{P}(\phi )\bigr]_{ij}=s_{ij}\ (e^{i\theta}e^{-i\phi}+1)\ \ \ \ \ \ \ \ (i\ne j,\ s_{ij}\ne0)  \label{nebb}.
\end{numcases}
By (\ref{neaa}) and (\ref{nebb}), the positivity of a matrix (\ref{dirww}) implies
the inequality
\begin{eqnarray}
P(\phi )\biggl[a_{i}a_{j}P(\phi )-\displaystyle\frac{\vert s_{ij}\vert ^{2}}{\vert\alpha\vert}\biggr ] \geq 0\ \ \ (s_{ij}\ne0)\label{dirxx}
\end{eqnarray}
for all $\phi$. However, there are some $\phi$ 
satisfying $0 < a_{i}a_{j}P(\phi ) < \vert s_{ij}\vert ^{2}/\vert\alpha\vert $.
Therefore the possibility of an off-diagonal element
in the form (\ref{dirpp}) is excluded
in linear, hermitian-preserving, and positive maps   
$\Lambda _{12}^{P}$, $\Lambda _{1}^{P}$, $\Lambda _{2}^{P}$.

It remains to consider an non-zero off-diagonal element $\bigl[\Lambda _{1}^{P}(\phi )\bigr]_{ij}\ (i\ne j)$
in the form (\ref{dirqq}) on the condition of $f_{ij}\ne 0$.
Then the hermiticity of a matrix (\ref{dirww}) implies
\begin{eqnarray}
\bigl[\Lambda _{1}^{P}(\phi )\bigr]_{ji}=\bigl[\Lambda _{1}^{P}(\phi )\bigr]^{\ast}_{ij}=s_{ij}^{\ast}r e^{-i\theta}f_{ij}^{\ast}\ (e^{i\phi}+f_{ij}^{\ast -1})(r^ {-1}e^{i\theta}e^{-i\phi}+1)\ \ \ \ (f_{ij}\ne 0,\ s_{ij}\ne0)  \label{necc}
\end{eqnarray}
This decomposition form $\bigl[\Lambda _{1}^{P}(\phi )\bigr]_{ji}$ is
realized in ($j, i$)-element of the form (\ref{dirqq}).
Then, there are two possible solutions 
\begin{eqnarray}
f_{ij}=r^ {-1}e^{i\theta}  \label{diraaa}
\end{eqnarray}
or
\begin{eqnarray}
r e^{i\theta}=r^ {-1}e^{i\theta} \label{dirddd}
\end{eqnarray}
because of the degree of freedom of interchanging $f_{ij}$ with $r e^{i\theta}$ in (\ref{dirqq}).
One immediately notices that the former solution (\ref{diraaa}) leads to
\begin{eqnarray}
\bigl[\Lambda _{1}^{P}(\phi )\bigr]_{ij}=\displaystyle\frac{s_{ij}e^{i\theta}}{\vert\alpha\vert}\ P(\phi)\ \ \ \ \ (i\ne j,\ s_{ij}\ne0).\label{dirccc}
\end{eqnarray}

On the other hand, the latter solution (\ref{dirddd}) implies $r=1$, therefore
the diagonal and the off-diagonal elements in this case become as follows:
\begin{numcases}{}
\bigl[\Lambda _{1}^{P}(\phi )\bigr]_{ii}=a_{i}\vert \alpha \vert e^{-i\theta}(e^{i\phi}+e^{i\theta})(e^{i\theta}e^{-i\phi}+1) \label{nedd}\\
\bigl[\Lambda _{1}^{P}(\phi )\bigr]_{ij}=s_{ij}(e^{i\phi}+e^{i\theta})(f_{ij}e^{-i\phi}+1)\ \ \ \ \ \ \ \ \ (f_{ij}\ne 0,\ i\ne j,\ s_{ij}\ne0). \label{neee}
\end{numcases}
From the positivity of a matrix (\ref{dirww}) constructed 
by the elements (\ref{nedd}) and (\ref{neee}), one easily obtains a condition
\begin{eqnarray}
f_{ij}=e^{i\theta}. \label{dirnnn}
\end{eqnarray}
This condition means that the off-diagonal element 
$\bigl[\Lambda _{1}^{P}(\phi )\bigr]_{ij}\ (i\ne j)$ in (\ref{neee})
also satisfies (\ref{dirccc}).

Eventually, the additional condition ``positivity'' implies that all the diagonal elements 
of a matrix $\Lambda _{1}^{P}(\phi )$ 
in the form (\ref{dirtt}), and the only allowed form of non-zero off-diagonal elements is given by 
(\ref{dirccc}). 
Thus a matrix $\Lambda _{1}^{P}(\phi )$ in Case 2 becomes
\begin{eqnarray}
\Lambda _{1}^{P}(\phi)= P(\phi)\Gamma_{1},  \label{dirmx}
\end{eqnarray} 
where $\Gamma_{1}$ is a fixed (i.e., independent of $\phi$) unit-trace operator,
so {\it uncorrelated cloning of a phase-set of states for all $\phi$ is impossible
in linear, hermitian-preserving, and ``positive'' maps}.

\section{Discussion and conclusion}\label{det}

While it has been known that
the linearity and the hermitian-preservation of maps exclude
the possibility to uncorrelatedly clone a phase-set of states deterministically 
by D'Ariano et al. \cite{dade}, we have shown that there exists 
a linear and hermitian-preserving
map to
uncorrelatedly clone a phase-set of states (for all $\phi$) probabilistically.
We illustrated that such a possibility is due to the anomalous relation
inherent to probabilistic uncorrelated maps (\ref{nej}), and one could generically classify it
under three cases according to how to inlay two factors of $e^{i\phi}P(\phi)$
into $e^{i\phi}\bigl[\Lambda _{2}^{P}(\phi )\bigr]_{\mu\nu}$.
Then only one of those three cases (Case 2) leads to the possibility of uncorrelated cloning,
but that possibility has eventually excluded by
the additional assumption ``positivity'' to maps $\Lambda _{12}^{P}$,
$\Lambda _{1}^{P}$, $\Lambda _{2}^{P}$.

In conclusion, we have proved that the condition of positivity in addition to
the linearity and the hermitian-preservation completely excludes all the possible maps
to uncorrelatedly clone a phase-set of states for all $\phi$ probabilistically.
Since quantum operations satisfy the linearity, the hermitian-preservation, and the positivity
at least, our result means that
it is impossible to uncorrelatedly clone a phase-set of input states for all $\phi$
probabilistically in quantum mechanics.

The positivity, which is used to derive our impossibility,
is essential to
guarantee the so-called ``probability interpretation'' in output operators
in quantum operations. 
While, from that reason,
the positivity is often assumed {\it a priori} in quantum mechanics, as a result, 
it tends to be overshadowed by the linearity \cite{bookb, bookc, gi, sbg}.
Furthermore, it seems intuitively that no difference occurs in the qualitative aspect of
quantum mechanics by whether one imposes the positivity or not
because quantum mechanics is considered to be founded on
``linear'' algebra belonged to by the other additional conditions such as
the hermitian-preservation, the positivity, and the complete positivity.
But our result of the impossibility of uncorrelated cloning
explicitly insists that the condition of positivity makes 
the qualitative difference in time evolution, namely, whether probabilistic 
uncorrelated cloning is possible or not.

On the other hand, conversely speaking, our derivation as to the impossibility by means of the
positivity also implies that
if one allows the existence of
{\it negative probabilities} which are induced in output operators,
it becomes possible to uncorrelatedly clone 
a phase-set of states probabilistically.
Since this fact predicts one of the consequences by extending the framework of quantum mechanics,
our result may become worth in the actively researched area recently called the
{\it generalized probabilistic theory} \cite{hardy, barre, inca} where one tries to
clarify the uniqueness of mathematical structure of quantum mechanics 
and to derive quantum mechanics from some fundamental physical principles.

\section*{Acknowledgements}
I thank S. Ishizaka, N. Hatakenaka, N. Ogita, T. Inagaki, and T. Ikuto for discussion and encouragement.

\appendix
\section*{Appendix}
\renewcommand{\thesection}{A}
A polynomial of first order of $e^{\pm i\phi}$ is written as
\begin{eqnarray}
W(e^{\pm i\phi} )=a e^{i\phi}+b e^{-i\phi}+c\ \ \ \ \ \ \ \ \ \ \ (a, b, c \in \mathbb{C}).  \label{dirlx}
\end{eqnarray}
Defining $x\equiv e^{i\phi}$ for the convenience of notation, the equation (\ref{dirlx}) becomes
\begin{eqnarray}
W(x )=a x+b x^{\ast}+c. \label{dirm}
\end{eqnarray}
If one multiplies both sides of the above equation (\ref{dirm}) by $x$,  one has 
\begin{eqnarray}
xW(x )=a x^{2}+c x+b. \label{dirn}
\end{eqnarray}
When $a\ne 0$ in (\ref{dirn}), 
$xW(x )$ is a polynomial of second order of variables $x$.
Then, by the ``unique polynomial factorization theorem''\footnote{This is the well-known classical theorem in algebra. See, for example, \cite{bookj}.},  
{\it a polynomial $xW(x )$ can be decomposed as 
\begin{eqnarray}
xW(x )=a (x+w_{0})(x+w_{1}) \ \ \ \ \ \ \ \ \ \ \ \ (a\ne 0,\ \ w_{0}, w_{1}\in \mathbb{C}), \label{diro}
\end{eqnarray}
where $w_{0}$ and $w_{1}$ are uniquely determined up to a permutation} \cite{bookj}.
By multiplying both sides of (\ref{diro}) by $x^{\ast}$,
one can again obtain
\begin{eqnarray}
W(x )=a (x+w_{0})(w_{1}x^{\ast}+1)\ \ \ \ \ \ \ \ \ \ \ \ (a\ne 0).  \label{dirp}
\end{eqnarray}
In (\ref{dirp}) as well as (\ref{diro}), one should note that $w_{0}$ and $w_{1}$ are uniquely determined up to a permutation
because if $W(x )$ were also written as
\begin{eqnarray}
W(x )=a (x+w'_{0})(w'_{1}x^{\ast}+1)  \label{dirq}
\end{eqnarray}
with the other pair of complex numbers $\lbrace w'_{0}, w'_{1}\rbrace$ different from 
the pair $\lbrace w_{0}, w_{1}\rbrace$, one has
\begin{eqnarray}
xW(x )=a (x+w'_{0})(x+w'_{1})  \label{dirr}
\end{eqnarray}
by multiplying both sides of (\ref{dirq}) by $x$.
However the difference between (\ref{dirr})
and (\ref{diro}) contradicts the uniqueness of polynomial factorization \cite{bookj}.
Therefore the pair $\lbrace w'_{0}, w'_{1}\rbrace$ must coincide with
$\lbrace w_{0}, w_{1}\rbrace$.

When $a=0$ in (\ref{dirm}), one can also confirm that $W(x )$ is uniquely written as
\begin{eqnarray}
W(x )=c(wx^{\ast}+1) \ \ \ \ \ \ \ \ \ \ (w\in \mathbb{C},\  c\ne 0) \label{dirs}
\end{eqnarray}
or
\begin{eqnarray}
W(x )=bx^{\ast} \ \ \ \ \ \ \ \ \ \ (c=0) \label{dirt}
\end{eqnarray}
with the same discussion above.

In the present paper, we call the forms of $W(x )$ appeared in the right-hand sides of
(\ref{dirp}), (\ref{dirs}), and (\ref{dirt}) the {\it decomposition forms} of a polynomial of first order of $e^{\pm i\phi}$.

\end{document}